\documentclass[graybox]{svmult}
\usepackage[mathscr]{eucal}
\newcommand{\Reals}{{\mathbb R}}         
\newcommand{\eps}{\varepsilon}
\newcommand{\NatNum}{\mathbb{N}}

\newcommand{\Pdd}{\mbox{$\partial$ \hspace{-1.1 em} $/$}}
\renewcommand{\L}{{\mathcal{L}}}

\newcommand{\la}{\langle}
\newcommand{\ra}{\rangle}
\newcommand{\Lin}{\text{\rm{L}}}

\newcommand{\R}{{\mathord{\mathbb R}}}
\newcommand{\tr}{\mbox{tr}}
\newcommand{\F}{\mathscr{F}}

\usepackage{type1cm}        
\usepackage{makeidx}         
\usepackage{graphicx}        
\usepackage{multicol}        
\usepackage[bottom]{footmisc}

\usepackage{newtxtext}       %
\usepackage{newtxmath}       

\makeindex             

\begin{document}

\title*{Proposal 42 - A new Storyline for the Universe based on the Causal Fermion Systems Framework}
\author{Claudio F. Paganini}
\institute{Claudio F. Paganini \at School of Mathematics, Monash University, 9 Rainforest Walk, 3800 Clayton, Victoria, Australia \email{claudio.paganini@monash.edu}
}
%
%
\maketitle

\abstract{Based on preliminary results from the Causal Fermion Systems framework regarding the matter-antimatter asymmetry in the universe, I propose a novel story line for the universe that would, if correct, resolve a number of problems in cosmology. First and foremost, the here-presented arguments suggest to identify cold dark matter as third generation (anti-)neutrino mass-eigenstates $\nu_3$. Furthermore, the proposal suggests a new look at the problem of initial conditions. Last but not least, the proposal also provides a new angle on the cosmological constant. }

\section{Introduction}
\label{42:sec:intro}
To this day we lack both a theory of quantum gravity and a fundamental theory of physics that manages to unify general relativity and the standard model of particle physics. Various candidates for quantum gravity/unified theory such as String Theory, Loop Quantum Gravity, Asymptotic Safety, Non-Commutative Geometry... have either failed to put forward falsifiable predictions or have not even succeeded in obtaining a fully consistent reproduction of our current models in an appropriate limit. There are some other approaches of which I want to highlight the work of Padmanabhan \cite{padmanabhan2017cosmic, padmanabhan2014cosmological, padmanabhan2012physical, padmanabhan2017we, padmanabhan2016atoms, padmanabhan2014general, padmanabhan2016exploring, padmanabhan2017atoms, padmanabhan2015gravity}. They arrive at a relationship between the CMB fluctuations and today's value of the cosmological constant. That series of work was started with \cite{padmanabhan2012physical}, for my considerations here it were especially the ideas in \cite{padmanabhan2017we} that served as an inspiration. A lot of the ideas that I will discuss in the present work are directly inspired and thus as far as I can judge compatible with the work of Padmanabhan, however and that is an important remark, they are not reliant on those ideas. I will point out, where appropriate, how the picture painted in Padmanabhans work can be fit into what I will present here. Despite some of the ideas being conceptionally close, they often come with a slight twist, at least at first sight there doesn't seem to be a neat one-to-one correspondence in the pictures (at least not in their underlying motivations). A detailed investigation of how the two pictures relate will however be postponed to later work.\\ 
The present work is based on ideas originating from the Causal Fermion Systems (CFS) framework, a novel approach to unification of the Standard Model of particle physics and General Releativity. A summary of the current state of knowledge on the framework can be found in \cite{finster1542continuum} for a proper introduction geared towards physicists see \cite{finster2015causal,dicefinster} for a introduction geared more towards mathematicians see \cite{finster2016causal}. I will include an introductory section to the CFS framework to make this paper sufficiently self contained. In my paper I will discuss how the CFS framework can give us a new look at the universe and how it might resolve some of the fundamental outstanding problems, and provide a new angle on others. As the ideas are heavily reliant on the CFS framework I will include a discussion on open questions in the framework and the physical interpretation, to the best of my current understanding.\\
The ideas put forward in the present paper are based on preliminary results regarding the matter anti-matter asymmetry in the universe, in the context of the CFS framework \cite{asymmetry}. If the here presented picture is correct, we could identify cold dark matter as third generation (anti-)neutrino mass-eigenstates. The approach, if correct, would not only explain the origin of the matter-/anti-matter asymmetry, but the origin of all matter and radiation. \\
Let me emphasize the novel part of the content here: The core of this paper is a new story line of the universe based on a new approach to unification. The story line does not only address a number of conceptional issues in today's models but also arrives at a prediction for the composition of Cold Dark Matter inside the Standard Model. \\
Note that most pieces of the story line for the universe presented here, are not entirely new. In fact most of the common pictures (inflation, big bang...) partially fit into the picture presented here. What is novel here is the fact that they are put together in a slightly different and conceptionally more consistent way.\\
The modifications of the Friedmann equations that are introduced have at most a weak theoretical motivation, however in interplay with the CFS mechanism for matter creation they lead to an instability mechanism that can induce a transition from a high $\Lambda$ phase to a low $\Lambda$ phase. According to \cite{padmanabhan2017we} such an instability could explain the arrow of time.   

\subsection*{Overview of the paper}
\label{42:sec:overview}
The paper is organized in the following way. In section \ref{42:sec:cfs} I will collect the fundamental notations and ideas of the CFS framework. In section \ref{42:sec:lambda} I will present a possible modification to the Friedmann equations that will facilitate the desired instability mechanism. In section \ref{42:sec:dgc} I will discuss a slightly different modification that, on universal scales, has a equivalent effect but might be better motivated by the CFS framework. In section \ref{42:sec:asym} I discuss the preliminary results regarding the matter and radiation creation mechanism derived in \cite{asymmetry}. In section \ref{42:sec:reheat} I will discuss how the matter creation mechanism leads to an uncertainty in the reheating time. In section \ref{42:sec:story} I will discuss how the pieces fit together to form a coherent story line for the universe. Finally in section \ref{42:sec:cfsdisc} I will discuss some of the open questions regarding the CFS framework. 

\section{Causal Fermion Systems} \label{42:sec:cfs}
As I assume most readers to be unfamiliar with the CFS framework I will here give a quick overview of the most important concepts and how they relate to well known structures. This section will be brief. For a proper introduction geared towards physicists see \cite{finster2015causal,dicefinster} for a introduction geared more towards mathematicians see \cite{finster2016causal}. For a overview of the state of the art of the CFS framework see \cite{finster1542continuum}. All the material in this section can be found in these sources. In section \ref{42:sec:cfsdisc} I will come back to the CFS framework and discuss some questions regarding the interpretation and open questions associated with it. 
\begin{remark}
When reading into the CFS framework for the first time, it is advisable to the reader to take all the concepts she knows from well established theories, put them in a box, and only get them out again, once she understands the abstract structures in themselves and is ready to start investigating the relations of concepts in the CFS framework to established theories.
\end{remark} 
The abstract definition of a CFS consists of three objects: a Hilbert space $\mathscr{H}$, a suitably chosen subset $\F$ of the linear operators on the Hilbert space $\Lin(\mathscr{H})$ and a measure $\rho$ that lives on $\F$.

\begin{definition}[Causal fermion system]\label{42:def:cfs}
Let~${(\mathscr{H}, \la .|. \ra_\mathscr{H})}$ be a Hilbert space. \\
Given a parameter {$n \in \NatNum$} (``{spin dimension}'') we set \\
${\F} := \Big\{ x \in \Lin(\mathscr{H}) \text{ with the properties:}$
\begin{itemize}
\item[] \hspace*{2cm} $x$ is {self-adjoint} and has {finite rank}
\item[] \hspace*{2cm} $x$ has {at most $n$ positive}
\item[] \hspace*{2.3cm} and~{at most~$n$ negative eigenvalues} $\Big\}$
\end{itemize}
and {$\rho$} a measure on~$\F$ (``{universal measure}'')
\end{definition}
From now on when I talk about a CFS I will always refer to a triple ${(\rho, \F, \mathscr{H})}$. If we have a CFS we can simply define a spacetime to be given by ${M:= \text{supp} \rho}$. With that definition we get that spacetime points are linear operators on $\mathscr{H}$\footnote{I will come back to that in section \ref{42:sec:cfsdisc}, where I will discuss a possible link between the CFS formalism and the Events, Trees, Histories (ETH) interpretation of quantum mechanics \cite{frohlich2016quest}.}.
In the following we want to introduce a causal structure on the spacetime $M$. For that we have to investigate the relationship between two points $x, y \in M$. The operator product $x \!\cdot\! y \in \Lin(\mathscr{H})$ in this case has non-zero, in general complex,  eigenvalues~${\lambda^{xy}_1, \ldots, \lambda^{xy}_{2n}}$. It is important to remark that $\F$ is not a group under operator composition. Hence for points~$x,y \in \F$ their product $ x \!\cdot\! y $ is not necessarily in $\F$ but it still has finite rank $\leq 2n$. Now using the eigenvalues of the operator product of $x,y$ we can define a causal structure on $\F$.
\begin{definition}[causal structure] \label{42:def:causalstructure}
The points~$x,y \in \F$ are called
\[ \left\{ \begin{array}{cll}
\text{{spacelike} separated} &\!&  {\mbox{if $|\lambda^{xy}_j|=|\lambda^{xy}_k|$
for all~$j,k=1,\ldots, 2n$}} \\[0.4em]
\text{{timelike} separated} &&{\mbox{if $\lambda^{xy}_1, \ldots, \lambda^{xy}_{2n}$ are
all real}} \\[0.2em]
&& \text{and $|\lambda^{xy}_j| \neq |\lambda^{xy}_k|$ for some~$j,k$} \\[0.4em]
\text{{lightlike} separated} && \text{otherwise}
\end{array} \right. \]
\end{definition}
In my opinion it is one of the core strength of the CFS framework that one can define consistent pairwise causality relations without appeal to any form of metric. For completeness I mention here that one can in fact write down a functional that can be used to define a time direction between two points. The following notation will be used
\[ x(\mathscr{H}) \subset \mathscr{H} \qquad \text{subspace of dimension $\leq 2n$} \]
\[ \pi_x : \mathscr{H} \rightarrow \mathscr{H} \qquad \text{orthogonal projection on~$x(\mathscr{H})$} \]
where $\pi_x$ is the projection operator on the image of the operator $x$. With this definition we can introduce the time direction functional
\begin{equation}
{\mathscr{C}} :  M \times M \rightarrow \R\:,\qquad
{\mathscr{C}}(x, y) := i \,\tr \big( y\,x \,\pi_y\, \pi_x - x\,y\,\pi_x \,\pi_y \big) 
\end{equation}
which leads us to define a time direction in the following way.
\begin{definition}[Time Direction]\label{42:def:order}
For timelike separated points~$x,y \in M$,
\[ \left\{ \begin{array}{cll}
\text{$y$ lies in the {future} of~$x$} &\!&  {\mbox{if ${\mathscr{C}}(x,y)>0$}} \\[0.4em]
\text{$y$ lies in the {past} of~$x$} && {\mbox{if ${\mathscr{C}}(x,y)<0$}}
\end{array} \right. \]
\end{definition}
In general the so defined time direction is not transitive. Therefore it is currently unclear whether this functional can be used to make any interesting global statements about causal ordering. See \cite{dappiaggi2018linearized} for recent progress with respect to that question. \\
Now we come to core of the CFS framework: The causal action principle. 
\begin{align}\label{42:eq:actionprinciple}
\text{{Lagrangian}} \quad {{\mathcal{L}}[A_{xy}]} &= \frac{1}{4n} \sum_{i,j=1}^{2n} \big(
|\lambda^{xy}_i| - |\lambda^{xy}_j| \big)^2 \;\geq\; 0 \\
\text{{Action}} \qquad
\quad {{\mathcal{S}}} &= \iint_{\F \times \F} {\mathcal{L}}[A_{xy}]\: d\rho(x)\: d\rho(y) \;\in\; [0, \infty]
\end{align}
A physical system is then given by a measure $\rho$ which minimizes the causal action ~${\mathcal{S}}$. To obtain the abstract CFS Eule--Lagrange equations we vary the action with respect to $\rho$ under the following constraints
\begin{align} 
\text{{volume constraint:}} && \rho(\F) = \text{const} \quad\;\; & \\
\text{{trace constraint:}} && \int_\F \tr(x)\: d\rho(x) = \text{const}& \\
\text{{boundedness constraint:}} &&
\iint_{\F \times \F} \left(\sum_{i=1}^{2n} |\lambda^{xy}_i|\right)^2\: d\rho(x)\, d\rho(y) &\leq {C}
\end{align}
\centerline{${C}$ determines regularization scale~${\varepsilon}$}
The volume constraint is rather natural and guarantees that the minimizing $\rho$ is non trivial. The other two constraints are of more technical nature. The technical definition of a minimizer is then given by
 \begin{definition}[Minimizer]
    $\rho$ is a minimizer if 
    $${{\mathcal{S}}}[\tilde\rho]-{{\mathcal{S}}}[\rho]\geq 0$$ for all $\tilde \rho$ with 
    $$|\tilde\rho-\rho|< \infty \qquad \text{ and } \qquad (\tilde\rho-\rho)\F=0$$.
    \end{definition}
Here, $|\tilde\rho-\rho|$ denotes the total variation. Using the notation 
\begin{equation}
    \ell(x) = \int_M \L(x,y)\: d\rho(y) 
\end{equation}
the abstract CFS Eule--Lagrange equations are then given by
\begin{lemma}[Euler-Lagrange equations] \label{42:lem:el}
\[ \ell|_M = \inf_\F \ell =: c\].
\end{lemma}
Hence for a measure to be a minimizer the function $\ell(x)$ needs to be constant for all $x\in  \text{supp} \rho $ and as a consequence the action $ {\mathcal{S}}$ is a multiple of the volume $ \rho(\F)$.
For most practical purposes the CFS Euler-Lagrange equations are linearized around a given minimizer. 

\subsection{Continuum Limit}\label{42:sec:continuum}
To make sense of the discussions to follow I need to briefly introduce how these structures relate to the classical notions in the continuum limit. Note again, what I will present here is at most a sketch, for a more detailed elaborations consult the sources mentioned before.\\
To represent Minkowsky space as a minimizer to the causal action principle we begin by looking at solutions to the free Dirac equation
\begin{equation}\label{42:eq:dirac}
    (i \gamma^k \partial_k - m) \,\psi = 0
\end{equation}
together with the usual scalar product 
\begin{equation}
    {\la \psi | \phi \ra} = \int_{t=\text{const}} (\overline{\psi} \gamma^0 \phi)(t, \vec{x})\, d\vec{x}
\end{equation}
where $\gamma^0$ is the Dirac matrix and $\overline{\psi} = \psi^\dagger\gamma^0$ is the adjoint spinor. One then chooses the Hilbert space to be all the negative energy solutions to equation \eqref{42:eq:dirac} and defines the local correlation operator via
\begin{equation}\label{42:eq:correlation}
    \la \psi | {F(x)} \phi \ra = -\overline{\psi(x)} \phi(x)  \qquad \forall \psi, \phi \in \mathscr{H}.
\end{equation}
Where one uses the fact that every bilinear form on a Hilbert space can be expressed via the scalar product and a linear operator. Note that the operator $F(x)$ is self-adjoint and has by definition finite rank $\leq 4$ and at most 2 positive and 2 negative eigenvalues. Therefore we have that $F(x)\in \F\subset L(\mathscr{H})$ for spin dimension $n=2$. The local correlation operator thus gives a mapping from the spacetime points into $\F$. We can then use the push-forward measure
\begin{equation}
    \rho(\Omega) := \int_{{F^{-1}(\Omega)}} d^4 x = \mu \big( {F^{-1}(\Omega)} \big)
\end{equation}
to obtain all the structures required by the definition of the CFS. \\
Now I wasn't quite honest here, as of course if we take the Hilbert space to be the solutions to \eqref{42:eq:dirac} with negative energy which are in $L^2(\Reals^3)$ then \eqref{42:eq:correlation} is generically ill-defined. Therefore to make the mapping well-defined we need to introduce a regularization
\begin{equation}
    \psi_\eps=R_\eps(\psi)
\end{equation}
where $R_\eps$ is the regularization operator - for example a convolution with a suitable kernel - such that $\psi_\epsilon$ is smooth and we have that
\begin{equation}
    \psi=\lim_{\eps\rightarrow 0}R_\eps(\psi). 
\end{equation}
This gives us the regularized local correlation operator 
\begin{equation}\label{42:eq:regcorrelation}
    \la \psi_\eps | {F_\eps(x)} \phi_\eps \ra = -\overline{\psi_\eps(x)} \phi_\eps(x)  \qquad \forall \psi, \phi \in \mathscr{H}.
\end{equation}
This is now a regularized CFS and the continuum limit is then defined as taking $\eps\rightarrow 0$ in a suitable sense.\\
It can be shown that in this limiting case the Dirac sea vacuum is a critical point of the causal action in a well-defined mathematical sense. In this sense, the Dirac sea drops out of the action, and only perturbations to the Dirac sea can be observed as matter and physical fields. In a way you can consider the Dirac sea to be the stage on which the world plays. An analogy I like to make is the following: You can think of the Dirac sea as the geography of a country while the actual physical world, i.e. the spacetime, is given by the travel time distance between any combination of two points in the country. It is obvious that the travel time distance between two places can change dramatically (for example when a new tunnel is opened) without the geography changing at all. A similar picture can be employed with respect to the Hilbert space, i.e. the Dirac sea in the CFS framework\footnote{Such analogies are of course always to be taken with the necessary grain of salt. They are intended to help the reader gain a rough intuition on the role different objects play in the framework.}. \\
I will only briefly mention one example here to establish a picture on how to think about some concepts in the context of the CFS framework. We now want to vary the vacuum minimizer. So let us introduce an external field as a vector potential $\mathcal{B}$ into the Dirac equation.
\begin{equation}
    {(i \Pdd + {{\mathcal{B}}} - m ) \psi = 0 \:. }
\end{equation}  
Now if we regularize and demand that the CFS Euler--Lagrange equation be satisfied in the limit $\eps\rightarrow 0$ in a suitable sense, we get as a result, that the vector potential has to satisfy Maxwell's equation. \\
Now this result can be viewed from two different sides: either the presence of the Maxwell field deforming the Dirac sea, or a collective deformation of the Dirac sea giving rise to a Maxwell field as an effective description and thus only the Dirac sea being fundamental and the Maxwell field being an emergent object. My favourite analogy at this point is the following: Suppose we would go for a walk at the beach, and for some weird reason I get interested in the shape of your feet. Now I will obtain the same information whether I ask you to show me your feet or whether I have a look at the collective behaviour of the sand grains where you just stepped a minute before.\\
It is important to stress that in the CFS framework electro-magnetic fields have no fundamental existence but are merely a simplified description of a phenomenon arising from the collective behaviour of all states in the Dirac sea. In fact the same is true for all bosonic fields in the CFS framework (including gravity). In summary the Dirac sea encodes all information about the physical world. 

\section{Dynamical 
$\Lambda$}
\label{42:sec:lambda}
\begin{figure}[h!]
    \centering
    \includegraphics[width=\textwidth]{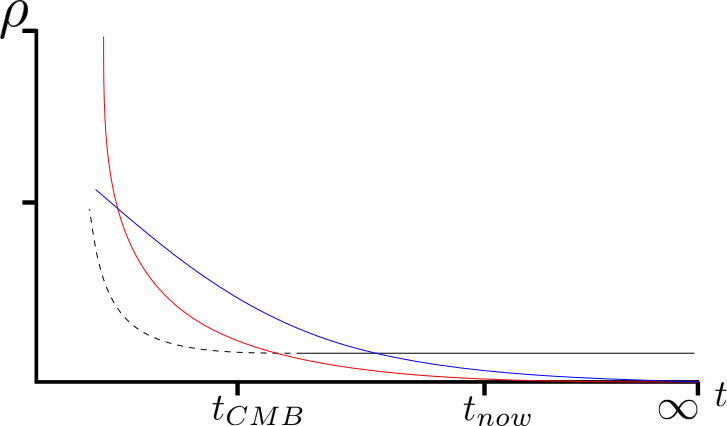}
    \caption{The graph shows a sketch of the expected future development in "time" (horizontal axis) of the energy densities (vertical axis) associated with radiation (red), matter (blue) and $\Lambda$ (black) under the Friedman equations with a running vacuum \eqref{42:eq:rchlambda} model. Significant modification to the value of $\Lambda$ are only expected in the very early universe. The standard sequence of having a radiation dominated phase that would go over to a matter dominated phase and then ending in a $\Lambda$ dominated phase is not affected. I used $t_{now}$ to roughly indicate where in the evolution of the universe we are today and $t_{CMB}$ to roughly indicate where the CMB decoupled in this picture.  }
    \label{42:fig:conjecture}
\end{figure}

Running or decaying vacuum models have been studied in recent years, see for example \cite{santos2018cosmological,coleman1980gravitational,polyakov2010decay,rajantie2017standard, freese1987cosmology,alcaniz2012cosmological,lima2013expansion} and some are indeed compatible with recent Planck data \cite{santos2018cosmological}. Also the set of ideas put forward by Padmanabhan \cite{padmanabhan2017cosmic, padmanabhan2014cosmological, padmanabhan2012physical, padmanabhan2017we, padmanabhan2016atoms, padmanabhan2014general, padmanabhan2016exploring, padmanabhan2017atoms, padmanabhan2015gravity} are based on the idea of a transition between two "cosmological energy levels", i.e. a high $\Lambda$ phase and a low $\Lambda$ phase, however to my best knowledge a proper mechanism describing the why and the how of this shift in "cosmological energy levels" seems to be absent from his work. It is the idea of a transition between "cosmological energy levels" that I will rely on in this paper. For the time being I will assign the mechanism to some sort of running vacuum model. At the current stage I have only a very weak theoretical motivation\footnote{The weak theoretical motivation mentioned here comes from my limited understanding of the CFS framework and how the objects therein might be interpreted. In section \ref{42:sec:dgc} I will discuss a different point of view which is superficially equivalent but not so commonly considered. This point of view might be more readily compatible with the CFS framework and seems to be favorable with respect to conceptional issues as well.} as to why I would want to consider the particular modification of the Friedman equation other than the fact that it provides the desired mechanism I need in section \ref{42:sec:story}. One might ask whether it makes sense to modify gravity on the level of the Friedmann equations. This can indeed make sense if the modification is an emergent phenomenon from the global behaviour of the universe in some underlying framework. In this case modifying the Friedman equation can indeed make sense, as the FRW coordinate system is special, as it can actually be determined by observations in the universe, see \cite{padmanabhan2017we} for a discussion. Therefore emergent phenomena can indeed enter on this level in the formulation of gravity.  \\
For the sake of the argument I will assume here the simplest possible running/decaying vacuum model. Hence we take the Friedmann equation
\begin{equation}\label{42:eq:friedmann}
    \frac{H^2}{H_0^2} = \Omega_{R} a^{-4} + \Omega_{M} a^{-3} +  \Omega_{\Lambda}
\end{equation}
where $H$ is the Hubble expansion rate, $\Omega_{R}$ the radiation density,  $\Omega_{M}$ the matter density and $\Omega_{\Lambda}$ the energy density related to the cosmological constant. I now impose that the dynamics of $\Lambda$ are given as a function of $H$ in a similar fashion to the modifications in \cite{santos2018cosmological}
\begin{equation} \label{42:eq:hlambda}
  \Lambda=3 H^2.  
\end{equation}
 Note that I leave the Friedmann equations unchanged beyond the fact that I allow for $\Lambda$ to depend on the Hubble expansion rate $H$. In the de Sitter Universe, hence when $\Omega_{R}=\Omega_{M}=0$ we have 
\begin{equation}
     \mathrm{d}s^2 = - \mathrm{d}t^2 + {a(t)}^2 \mathrm{d}{\sigma}^2
\end{equation}
with
\begin{equation}
    a(t) = e^{Ht}
\end{equation}
where $\sigma$ is the flat metric in $\Reals^3$ and $a(t)$ is the scale function. We have that the relation \eqref{42:eq:hlambda} is satisfied. If we write de Sitter space in stationary coordinates instead
\begin{equation}
    ds^2 = -\left(1-\frac{\Lambda r^2}{3}\right)dt^2 + \left(1-\frac{\Lambda r^2}{3}\right)^{-1}dr^2 + r^2 d\Omega_{n-2}^2.
\end{equation}
we get that the cosmological horizon $r_{CH}$ is located at
\begin{equation}\label{42:eq:rch}
    r_{CH}=\sqrt\frac{3}{\Lambda}= \frac{1}{H}.
\end{equation}
In a slight abuse of notation i will in the following refer to de-Sitter as a "stationary" solution to the system of equations given by \eqref{42:eq:friedmann} and \eqref{42:eq:hlambda}\footnote{I consider this to be a sensible notation as for any two times $t_a$ and $t_b$ the hypersurfaces can be mapped into each other isometrically hence there is no change that could be observed by an outside observer. In the absence of matter or other perturbations the notion that de-Sitter space is exponentially expanding is absolutely meaning less. It only acquires meaning once you put test matter (or any group of test bodies for that matter) there, tracing out a geodesic foliation with their proper time functions.}.
In the following I will use the radius of the causally interacting region $r_{CH}$
\begin{equation}\label{42:eq:rchlambda}
\Lambda=\frac{3}{r_{CH}^2}
\end{equation}
to parametrize the dynamics of $\Lambda$ instead of \eqref{42:eq:hlambda} as it is more natural to the way of thinking about the universe I want to introduce later on. If we consider the relation as functions in terms of the scale function $a(t)$ then we get the following derivative
\begin{equation}\label{42:eq:lambdader}
\frac{d\Lambda}{d a}=-\frac{6}{r_{CH}^3}\frac{d r_{CH}}{da}.
\end{equation}
One sees immediately that the change in $\Lambda$ is more dramatic in the early universe when $r_{CH}$ is small and $\frac{d r_{CH}}{da}$ is non vanishing due to the fact that the universe is either matter or radiation dominated. In the late universe on the other hand $r_{CH}$ is large and particularly in the $\Lambda$ dominated era $\frac{d r_{CH}}{da}$ is close to zero hence the change in $\Lambda$ should be small. This gives an intuition for why the compatibility of this sort of running vacuum models with Planck data reported in \cite{santos2018cosmological} comes at no surprise.\\
On the conceptional level it would be necessary to check the following conjecture.
\begin{conjecture}\label{42:con:tods}
Let $\rho_R(a(t))|_0\geq0$, $\rho_M(a(t))|_0\geq0$ and $\Lambda(a(t))|_0>0$ be initial data at some time $t=t_0$ (or equivalently at some value of the expansion parameter $a=a_0$) to the system of equations consisting of \eqref{42:eq:friedmann} and \eqref{42:eq:rchlambda}. Then we have that 
\begin{equation}
    \lim_{t\rightarrow\infty}\rho_R=\lim_{t\rightarrow\infty}\rho_M=0 \qquad \text{ and }\qquad \lim_{t\rightarrow\infty}\Lambda(a(t))=\Lambda_\infty>0.
\end{equation}
\end{conjecture}
Here I denote with $\rho_M$ the physical energy density of matter and with $\rho_R$ the physical energy density of radiation in a homogenous universe. In colloquial terms that conjecture states that if the universe evolves under equations \eqref{42:eq:friedmann} and \eqref{42:eq:rchlambda} it will asymptotically approach to a de-Sitter universe. In figure \ref{42:fig:conjecture} you find a sketch of this expected behaviour of the various densities. 

\subsubsection{Thermodynamic Interpretation}
 This section is intended to show that the above modification of the equations could indeed have a deeper theoretical motivation. It is however quite speculative and not fully expanded. Thus this section can be ignored by anybody just interested in the larger picture. \\
For black holes one can define an entropy via the surface area $A$ in dimensionless form
\begin{equation}\label{42:eq:bhentropy}
    S_{BH}=\frac{A_{hor}}{4 L_P^2}=\frac{c^3A_{hor}}{4G\hbar}
\end{equation}
where $L_P$ stands for the Planck length, G for Newton's constant, $\hbar$ for Planck's constant and $c$ is the speed of light. Analogous we can define a cosmological entropy for de-Sitter space by
\begin{equation}\label{42:eq:cosmentropy}
    S_{C}=\frac{A_{CH}}{4 L_p^2}=\frac{c^3A_{CH}}{4G\hbar}
\end{equation}
where the surface area is given by $4\pi r_{CH}^2$ and hence with equation \eqref{42:eq:rch}
\begin{equation}
    A_{CH}= \frac{12\pi}{\Lambda}
\end{equation}
and thus
\begin{equation}
    S_C=\frac{3\pi}{L_P^2 \Lambda}= \frac{3\pi c^3}{\hbar G \Lambda}.
\end{equation}
With this expression one could in fact use 
\begin{equation}\label{42:eq:entropylambda}
    \Lambda=\frac{3\pi}{L_P^2 S_C}
\end{equation}
as a definition for the running vacuum with a slight further modification
\begin{equation}\label{42:eq:fullentropylambda}
    \Lambda=\frac{3\pi}{L_P^2 (S_C+S_M+S_R)}.
\end{equation}
Here $S_M$ stands for the entropy associated with matter and $S_R$ stands for the entropy associated with radiation inside the interacting region.
Equation \eqref{42:eq:fullentropylambda} is a slight variation on \eqref{42:eq:entropylambda} which I will not get further into. Both modifications \eqref{42:eq:entropylambda} and \eqref{42:eq:fullentropylambda} satisfy the condition that if coupled with \eqref{42:eq:friedmann} the de-Sitter universe, i.e. de Sitter space is a stationary solution. In fact in this case \eqref{42:eq:entropylambda} and \eqref{42:eq:fullentropylambda} are identical because $S_R=S_M=0$\footnote{In fact for the considerations later on the only thing we need is for the modification to allow for de-Sitter space as a solution and for the instability mechanism to arise in the right form, all without modifying the late time behaviour too much. The choice made here is simply to demonstrate that such modifications in fact exist and \cite{santos2018cosmological} shows that the compatibility with observations can be arranged.}.\\
It is interesting to note that with this interpretation entropy would play an active role in the evolution. In a sense given conjecture \ref{42:con:tods} is true and we interpret the running vacuum as suggested by the form of equation \eqref{42:eq:entropylambda}, we would have that the presence of entropy itself leads to the following conjecture for the second law of thermodynamics 
\begin{conjecture}\label{42:con:entropy}
Suppose the universe evolves according to equation \eqref{42:eq:friedmann} coupled to \eqref{42:eq:entropylambda} than we have that 
\begin{equation}
    \frac{d(S_C+S_M+S_R)}{dt}\geq0
\end{equation}
holds on universal scales, and 
\begin{equation}
    \frac{d(S_C+S_M+S_R)}{dt}=0
\end{equation}
is only true in the stationary case of de-Sitter space where $\rho_M=\rho_R=0$ and hence $S_M=S_R=0$.
\end{conjecture}
Now how does this all connect to the CFS framework? I have to be honest that from this point of view it is extremely vague and based on the following open question:
\begin{question}{Thermodynamic Interpretation of CFS}
Despite all the results that have come from the CFS framework (see \cite{finster1542continuum} for an overview) and despite how well it fits in the limitting case with the well established theories, its major weakness, in my opinion, is the absence of a clear physical interpretation of the variation principle \ref{42:eq:actionprinciple} that lies at the heart of the formalism. \\
One possible avenue towards such a physical interpretation is looking at the mechanism behind the continuum limit where spacetime points are represented by local correlation operators \ref{42:eq:correlation}. Thus the minimization of the action principle tries to reduce the correlation in the spacetime which one could try to interpret in an entropic sense as trying to maximize the amount of disorder in the spacetime.\\ If the above scenario with a dynamical cosmological constant can be realized as a Causal Fermion System, then conjecture \ref{42:con:entropy}, if true, would support this interpretation. 
\end{question}
In the following section I will present a short discussion of an alternative modification of the Friedmann equations that superficially seems very similar to what I presented in this section but seems to fit more naturally in the CFS framework. 
\subsection{Dynamical Gravitational Coupling} \label{42:sec:dgc}
This section mainly serves to discuss some very general points about an alternative modification of the Friedmann equations. I include it here because at least intuitively it seems to lead to a very similar evolutionary behaviour of the universe as in section \ref{42:sec:lambda}\footnote{This claim has not been checked beyond the rough sketch of arguments that I will present here.} however it is more compatible with the CFS framework. For more details see upcoming work by Finster and R\"oken \cite{dgcfinster}. This section is not relevant for the understanding of the general picture that I will present in the following. I will usually refer to the modification in section \ref{42:sec:lambda} which have been more widely considered in the literature. However I will reference this section whenever the two modifications are conceptionally different.\\
If we look at the right hand side of the Friedmann equation \eqref{42:eq:friedmann} we see that 
$$\Omega_{R}a^{-4}+ \Omega_{M} a^{-3}  \propto G $$
and
$$\Omega_{\Lambda} \propto \Lambda.$$
Hence if we are only interested in the question whether the universe is matter, radiation or $\Lambda$ dominated, we can simply divide the whole equation by $G$ which gives us a rescaling of $H_0$ and the last term on the right hand side being
$$\Omega_{\Lambda} \propto \Lambda/G$$
Hence if instead of letting $G$ be constant and $\Lambda\propto H ^2$ we assume $\Lambda$ to be constant but the gravitational constant $G$ to change with $G\propto1/H^2$, we arrive at the same conclusion (without considerations of back reactions due to a possibly different evolution of $H$) for the question whether the universe is matter, radiation or $\Lambda$ dominated in a certain period. In the period between the CMB and now I would expect the difference between the two modifications to be small and hence it is only really the change of that ratio that is relevant for the global dynamics in that period. See for example \cite{jamil2009holographic} for a discussion of the compatibility of a dynamical gravitational constant with observations. In the period where one would expect a rapid change in either $\Lambda$ or $G$ in these modifications, the issue might be different. Of course I completely ignored here that by dividing the whole equation by $G$ assigning the dynamics to $G$ instead of $\Lambda$ would also modify the left hand side. In particular a small change in $G$ between the CMB time and now might easily fix the $H_0$ tension between local observations \cite{riess20162} and CMB measurements by the Planck collaboration \cite{planck2016astronomy}\footnote{This by itself is of course no feat as introducing an additional functional degree of freedom allows one to fit almost anything.}.\\
It is important to note, that also with this modification - a dynamical gravitational constant - de-Sitter space is a stationary solution to the Friedmann equations. This is simply due to the fact that in this case we have
\begin{equation}
    \rho_M=\rho_R=0
\end{equation}
and thus the value of $G$ plays absolutely no role. \\
This modification is much more natural to the CFS framework as we have $G\propto L_P^2$ and $L_P$ is essentially the regularization length scale $\eps$ mentioned in section \ref{42:sec:cfs}\footnote{Note, that this suggests, that for the evolution of the universe only the ratio of the $\Lambda$ and $L_P^2$ is relevant. This in a sense resolves the problem that there are two fundamental energy scales in the universe, as this ratio is dynamical in both scenarios. Hence only one of them sets a fundamental scale.}. The detailed dynamics of such a dynamical gravitational coupling are however dependent on the choice of regularization. This leads to the following open question.

\begin{question}{Correct Choice of Regularization}
For the CFS framework to make sense one needs to make a choice of a particular regularization. This leads of course to a whole array of questions on how to choose the correct regularization. It is important however to note that this dependence on the choice of regularization is not a fundamental problem, as long as we can choose it coherently for all pieces of the story line, as we do actually have a preferred coordinate system in the universe, see cf \cite{padmanabhan2017we}. Locally when one sets up a experiment in the lab, implicitly one prepares the experiment such that its local surrounding looks like a vacuum. Thereby one essentially chooses a local reference frame and its corresponding regularization when setting up the experiment.
\end{question}
Further we have the generic open question for this section
\begin{question}{Derivation of Correct Modification}
The modifications discussed here are only a few of many possibilities. The choice made here is just to illustrate a point. Eventually one needs to be able to derive an adequate modification on the level of the Friedmann equation or in a more general setting from fundamental principles. Otherwise one performs just phenomenology. The considerations in this paper might serve as a guideline for the search of the correct modification.
\end{question}

\section{Mechanism of Matter Creation} \label{42:sec:asym}
In the following section I will discuss some preliminary results from my upcoming work with Finster \cite{asymmetry}. It is important to note that many details are still not fully worked out, so anything I discuss here has to be taken with a grain of salt. Where the functional dependence is unclear I will indicate so by adding a $\#$ to the parameters.\\
The core objective of \cite{asymmetry} was to work out the origin of the matter-/anti-matter asymmetry in the context of the CFS framework. However I will argue here and in the following section that, provided the mechanism holds true, we achieve much more than that. In fact, the picture that emerges is that all matter and radiation in the universe could be a direct result of the matter-/anti-matter asymmetry. I will discuss the mechanism behind that observation in the present section and how we arrive from this considerations at a dark matter candidate in the next section \ref{42:sec:reheat}.\\
It is probably fair to mention that obtaining a matter-/anti-matter asymmetry from the CFS framework by itself is an interesting and important result. This result is not entirely surprising however, given that the CFS framework has a built-in asymmetry how it treats the matter and anti-matter states, by choosing the Dirac sea as the relevant Hilbert space. \\ 
Assuming the Dirac sea to be real and the regularization $\eps>0$ there is a finite number of states in the Dirac sea in a box or a finite density of states in the Dirac sea in $\Reals^{3+1}$. Now if you change the regularization or enlarge the box, you will get a change in the number of states required to fill the Dirac sea. If you do both simultaneously it can be arrange that the two effects cancel out, you get a dynamical gravitational constant but no change in particles. On the other hand, if things play our right you will essentially get a sea level shift, either up or down. As long as the effect is of the right order it does not matter whether we are in a situation with a sea level shift up, i.e. particles created with minimum kinetic energy, or down, i.e. anti-particles created with zero kinetic energy. This last bit is important to understand. The sea level shifting from the Dirac sea configuration will fill the particle or empty the anti-particle states starting from the one with zero kinetic energy successively filling/emptying the states with higher and higher kinetic energy. This observation will play a central role in the considerations of the next section.\\
The CFS Euler--Lagrange equation (CFS EL, Lemma \ref{42:lem:el}) play a crucial role in the mechanism of matter creation described in \cite{asymmetry}. Finster's work in \cite{finster1542continuum} shows that to leading order in $\eps$, the CFS EL reproduce the classical field equations. Now to obtain the matter/anti-matter asymmetry we need to consider next to leading order corrections to the CFS EL. Hence all effects will come with a pre-factor of the regularization length scale $\eps\propto L_P$. This suggests that the asymmetry can be seen as a quantum gravitational effect. The details will be presented in \cite{asymmetry}. The only things relevant for my present discussion are the following. The mechanism currently seems to obey the pattern that a Friedmann-Robertson-Walker spacetime to leading order in the CFS EL maintains particle number conservation. Only once we add perturbations and solve the CFS EL to first order we get a non-zero effect. 

\begin{align*}
    \text{\textbf{a)} Homogenous FRW + zeroth } & \text{order solution to the CFS EL}\\
    & \Downarrow\\
    \text{particle num} & \text{ber conservation}\\\\
    \text{\textbf{b)} Homogenous FRW + first } & \text{order solution to the CFS EL}\\
    & \Downarrow\\
    \text{particle num} & \text{ber conservation}\\\\
    \text{\textbf{c)} FRW+ perturbations + zeroth } & \text{order solution to the CFS EL}\\
    & \Downarrow\\
    \text{particle num} & \text{ber conservation}\\\\
    \text{\textbf{d)} FRW+ perturbations + first } & \text{order solution to the CFS EL}\\
    & \Downarrow\\
    \text{change in p} & \text{article number}
\end{align*}
So the matter/anti-matter asymmetry (or as I will argue in the next section, the existence of all matter and radiation) in the universe arises dynamically due to an interaction between vacuum fluctuations and next to leading order, i.e. quantum gravity, corrections to the field equations. The CFS formulation that leads to the standard model consists of 8 fermionic sectors, 6 for the quarks (for each colour of each typ one), one for the charged leptons and one for the neutrinos. Each sector contains exactly 3 generations i.e. the electron, the muon and the tau all live in the same sector. To get the correct results for the Standard Model out of the CFS framework, the violation of the chiral symmetry in the neutrino sector - i.e. that all observed neutrinos to this day have been lefthanded - has to be built in by hand.\\
We make the following assumptions: for every fermionic generation the naked mass of the quarks and the charged lepton are identical\footnote{This leaves the effective masses of the quarks and charged leptons to be explained.  Except for the tau, the effective masses of the quarks and their charged leptonic counterparts are within an order of magnitude. Recent calculations show that the binding energy in the strong force can account for almost all the mass in the neutrons and protons, see for example \cite{wilczek2012origins}. In that light I consider it plausible that the difference in effective masses within a generation can be accounted for  by self-interaction effects.}, but the naked mass of the neutrino is different. This assumption is crucial as I will explain below. Along the notation in \cite{finster1542continuum} I will use $\beta\in \{1,2,3\}$ to label the fermionic generations. So for every generation we have the naked mass of the quarks and leptons denoted by $m_\beta$ and the difference between $m_\beta$ and the naked mass of the neutrino denoted by $\Delta_{m_\beta}$. The density of particles created then obeys the following formula \cite{asymmetry}
\begin{equation}\label{42:eq:numberdensity}
    \frac{d\rho_{particles}}{dt}=\eps*N(m_\beta,\Delta_{m_\beta},\#)*(3 q_{1,\beta} + 3 q_{2,\beta} + l_\beta + \alpha (\Delta_{m_\beta}, \#) * \nu_{\beta})
\end{equation}
where $q_{1,\beta},q_{2,\beta}$ denote the two quarks in a generation, $l_\beta$ stands for the charged lepton in a generation and $\nu_{\beta}$ denotes a particle in the (anti-)neutrino mass-eigenstate of that generation.\footnote{Here I would like to emphasize that it is the mass-eigenstates that matter for the particle creation and not the eigenstates of the weak interaction. } The factor $\eps*N(m_\beta,\Delta_{m_\beta},\#)$ is the number density of particles created per volume per unit of time\footnote{"Time" here has to be considered a place holder for an appropriate parameter that tracks the history of the universe.} and $\alpha (\Delta_{m_\beta}, \#) $ is a dimension less function that describes how many neutrinos are created in relation to the quarks and charged leptons. 
I will now make some assumptions on the functions $N$ and $\alpha$ which are mostly based in preliminary results of our upcoming work \cite{asymmetry}. As mentioned in the introduction to this section I use the $\#$ symbol to indicate possible further functional dependencies which are unclear in the current state of the work on \cite{asymmetry}. I will here assume that $N$ also depends on the amplitude of the perturbations which I will denote with $\mathcal{A}$ as well as on the value of the cosmological constant $\Lambda$, hence $N=N(m_\beta,\Delta_{m_\beta},\mathcal{A}, \Lambda)$. For $\alpha$ I will consider no further dependencies, hence $\alpha=\alpha (\Delta_{m_\beta})$.\\
Preliminary results for \cite{asymmetry} suggest that for $\Delta_{m_\beta}=0$ we have $\alpha (0)=1$ and $N(m_\beta,0,\mathcal{A}, \Lambda)=0$. Therefore the mechanism crucially requires for the naked mass of the neutrinos to be different from the naked mass of the quarks and the charged leptons. In the following I will therefore assume that $\Delta_{m_\beta}\neq0$ and treat $\alpha$ as a free parameter which one can adjust at will. In principle it can also be negative which would mean that a corresponding number of anti-neutrinos is created.\\
For $N$ the preliminary calculations suggest that for $m_{\beta_1}<m_{\beta_2}$ we have that 
\begin{equation}\label{42:eq:g3matters}
    N(m_{\beta_1})\ll N(m_{\beta_2}).
\end{equation} I will further assume that $N$ is linear in $\mathcal{A}$ and that we have a strong suppression of the effect with declining $\Lambda$. Hence that for $\Lambda_1>\Lambda_2$ we have $N(\Lambda_1)\gg N(\Lambda_2)$\footnote{One way to motivate this is to look at the stationary holographic equipartition in \cite{padmanabhan2017atoms}. Where an equality between the gravitational degrees of freedom in the bulk and the boundary is stated for stationary spacetimes, hence $N_{bulk}=N_{boundary}$. With \eqref{42:eq:cosmentropy} this leaves $N_{bulk}\propto \Lambda^{-1}$. Together with equation \eqref{42:eq:rch} this leaved the density of degrees of freedom in a spacelike hypersurface to be $N/V\propto\sqrt{\Lambda}$ which decreases with decreasing $\Lambda$. It is conceivable that this could reduce the rate of particles created. However for the present work we will just impose this reduction as a condition.}.\\
For the time being the reason to choose these additional conditions is solemnly motivated by the fact that they resolve conceptional issues. Whether they are compatible with/can be derived from the fundamental framework remains to be seen. It is conceivable that the functional dependencies will turn out to be different and hence the story line presented in section \ref{42:sec:story} will need to be adapted.\\
In the next section I will discuss the consequence of equations \eqref{42:eq:numberdensity} and \eqref{42:eq:g3matters}. 


\section{Reheating Uncertainty} \label{42:sec:reheat}
From here on for the rest of the paper I will assume that the matter creation will happens instantaneously. Hence after the creation event we have the following matter density $\rho_{M, \beta g}$ for each generation of fermions
\begin{equation}\label{42:eq:instantnumberdensity}
    \rho_{M, \beta g}(t_0)=\eps*N(m_\beta,\Delta_{m_\beta},\#)*(3 q_{1,\beta} + 3 q_{2,\beta} + l_\beta + \alpha (\Delta_{m_\beta}, \#) * \nu_{\beta}).
\end{equation}
For now I will also assume that the kinetic energy at creation time is vanishing in comparison to the mass energy\footnote{Remember that in section \ref{42:sec:asym} I discussed the fact that the newly created particles fill the kinetic states starting from the one with the least momentum.}. First, I want to start this section with two thought experiments that will serve to put the later parts into context. 
\begin{remark}[Thought Experiment 1]
If an atom is coupled to an electro-magnetic field the probability for an atom to transition between two states is identical in both directions, i.e. for emission and absorption. So if I put an excited atom into a perfect cavity, it will decay, emit a photon and then sit around. The photon will bounce around and eventually be reabsorbed by the atom. You can play the movie forward and backward and everything is fine. Now suppose we play this game in the universe where we place the mirror to reflect the emitted photon 1 billion light years away. All technicalities aside, even if the photon manages to return to the atom it will never be reabsorbed because it has redshifted due to the expansion.The expansion of the universe therefore acts as a dispersion mechanism. Without which quantum mechanical systems would not tend towards the ground state even if coupled to electro-magnetic fields.
\end{remark}
The second thought experiment is more closely related to the actual process I will be discussing in this section. However the first one ties into it as well. The following thought experiment is based on the observation that "free" energy (i.e. energy not bounded as mass) has a different effect on the evolution of the universe whether it is bounded in a particle or present as radiation. This can be easily seen from equation \eqref{42:eq:friedmann}.
\begin{remark}[Thought Experiment 2]
Suppose you have a universe with one sort of atoms that have a rest mass of $m_1$ and one excited state that contains $E_2>>m_1c^2$ in energy. Now suppose all the atoms are excited in the beginning and they are entangled (so the excited states decay simultaneously) then the expansion history of the universe looks completely different depending on how long the excited state takes to decay. \\
Before the decay the expansion is matter dominated, after the decay the expansion is radiation dominated and eventually goes back to matter dominated.\\ 
So now we are in the curious situation where we have a universe for which we know the expansion history only in a probabilistic sense even though we treat gravity completely classical. 
\end{remark}
This second thought experiment brings us to one of the many uncertainties regarding the reheating process originating from the matter creation mechanism introduced in section \ref{42:sec:asym}. Now first I need to discuss how we arrive at a reheating scenario from the considerations in the previous section. Here the condition \eqref{42:eq:g3matters} plays a crucial role. We will assume that $m_3>m_2>m_1$ and therefore we get that $N(m_3)\gg N(m_2)\gg N(m_1)$ and thus for all practical purposes we can effectively assume $N(m_2)=N(m_1)=0$ and thus all matter being created in the third generation of fermions i.e.
\begin{equation}\label{42:eq:mattercreation}
    \rho_{M, 3g}(t_0)= \eps N(m_3, \Delta_{m_3}, \mathcal{A}, \Lambda) (3 t + 3 b + 1 \tau + \alpha (\Delta_{m_3}) \nu_3).
\end{equation}
This leaves us exactly in the scenario of the second thought experiment. For simplicity we will assume here an instant reheating at $t=t_{rh}\geq t_0$. In this context, reheating means a decay of the tops, bottoms and taus into first generation fermions plus an amount of leftover ``free'' energy, i.e. radiation
\begin{equation}
    \rho_{M, 3g}(t_{rh})=\frac{\rho(t_0) a(t_0)^3}{a(t_{rh})^3}= \rho_M(t_{rh}) + \rho_R(t_{rh}).
\end{equation}
Here $\rho_M(t_{rh})$ represents the energy density bounded in the mass after reheating, i.e. after the decay, and $\rho_R(t_{rh})$ represents the energy density set free as radiation in the course of the reheating process. Now it is clear that the actual reheating process in this scenario is a very complicated process and precise computations will require the insight of people more proficient in high energy physics than myself. So I will present only some very simple calculations here and I will have to leave the precise treatment to future research.\\
For a first approximation I will assume that in the initial state, all the energy is stored in the mass of the third generation fermions. That means that there are no composite third generations particles and hence no component in the energy density coming from the binding energy. Further I will assume that in the final state after the reheating there is no binding energy stored in the first generation quarks. In this case we would have 
\begin{align}
    \rho_R(t_{rh}) & \approx\eps N(m_3, \Delta_{m_3}, \mathcal{A}, \Lambda,t_{rh} ) (3 t + 3 b + 1 \tau)\\
    \rho_M(t_{rh}) & \approx \eps N(m_3, \Delta_{m_3}, \mathcal{A}, \Lambda,t_{rh} ) (3 u + 3 d + 1 e + \alpha (\Delta_{m_3}) \nu_3) + \rho_{\nu_\mu, \nu_e}
\end{align}
where $\rho_{\nu_\mu, \nu_e}$ represents the collection of all hot muon neutrinos and electron neutrinos that are created in the reheating process. If we ignore the uncertainty connected to that term for a moment we would get that
\begin{equation}
    \frac{\rho_R(t_{rh})}{\rho_M(t_{rh})}\approx 10^5
\end{equation}
To me it is at the moment absolutely unclear how much binding energy there is in the initial state before the reheating, however it is clear that the baryonic energy density in the universe is far greater than just the mass of the up and down quarks. From the simple calculations above we would get an equal number of neutrons and protons\footnote{If we take $\rho_M(t_{rh })  \propto \left(1p + 1n + 1 e +\alpha \nu_3\right) + \rho_{\nu_\mu, \nu_e}$ the ratio $ \frac{\rho_R(t_{rh})}{\rho_M(t_{rh})}\approx 10^2$ drops significantly. Adding binding energy to the third generation fermions prior to reheating would increase that ratio on the other hand.}. However we know from observations that the neutron to proton ratio at the neutrino decoupling time $t=t_{\nu }$ is close to $1/6$ and hence we have
\begin{equation}
    \rho_M(t_{\nu })  \approx \eps N(m_3, \Delta_{m_3}, \mathcal{A}, \Lambda,t_{\nu } ) \left(\frac{12}{7}m_p + \frac{2}{7}m_n + \frac{12}{7} m_e +\alpha m_{\nu_3}\right) + \rho_{\nu_\mu, \nu_e}
\end{equation}
Let us ignore $\rho_{\nu_\mu, \nu_e}$ for the moment. Further assume that neutrons and protons have the same mass. Given that for the moment we can treat $\alpha$ as a free parameter, in this scenario the third generation (anti-)neutrino mass-eigenstates make up all the Cold Dark Matter in the universe\footnote{A more precise formulation would be that this scenario is compatible with a universe where third generation (anti-)neutrino mass-eigenstates make up Cold Dark Matter. If other observations would in fact rule out third generation (anti-)neutrino mass-eigenstates this would unfortunately not suffice to kill the narrative presented here as we can - at present - simply adjust $\alpha$ to the bounds given by those experiments. In that case however the scenario presented here would of course fail to account for Cold Dark Matter, which would weaken its case. We hope to be able to constrain $\alpha$ from fundamental considerations in future research.}. This leaves us with 
\begin{equation}
    \frac{85\%}{15\%}=\frac{\rho_{CDM}}{\rho_B}= \frac{|\alpha| m_{\nu_3}}{2p}.
\end{equation}
The mass of the proton is roughly $940MeV$ the upper bound for the mass $m_{\nu_3}$ is roughly $15MeV$ which gives us a lover bound for $|\alpha|$
\begin{equation}
    |\alpha|\geq 710.
\end{equation}
Hence if the picture discussed here were to be true we should expect to find roughly 500 third generation (anti-)neutrino mass-eigenstates for every neutron and proton in the universe. I would be surprised if that abundance had no observational effect. I will come back to that in section \ref{42:sec:conclusion}\footnote{Note that for example such an abundance would render the total matter content in the universe having a strong chirality asymmetry.}. \\
If the picture presented in this section is true than the entire physical content of the universe is initially created in equal numbers of tops and bottoms, albeit the tops dominating due to their mass\footnote{If this picture is true and the universe was created by some sort of god, she certainly must have a good sense of humor.}.\\
For the moment I have to leave two minor and three major questions in this section open.
\begin{question}{Radiation to Mass Energy Ratio}
Deriving a precise ratio of the energy density of radiation and matter after the reheating phase might prove to be a very challenging problem. Note that if one manages to derive sharp relative values for $\rho_M(t_{rh})$ and $\rho_R(t_{rh})$ this would allow to fix $\rho_{eq}=\frac{\rho_M(a)^4}{\rho_R(a)^3}$ from first principles. This ratio plays a central role in the considerations of Padmanabhan \cite{padmanabhan2017we} and is in fact a time-independent, observable quantity in the universe. 
\end{question}
Connected to that is the first minor questions:
\begin{question}{Correct Neutrino decoupling}
From observations we get that the proton to neutron ratio is 6:1 at neutrino decoupling time. From the particles created in the beginning a decay strictly within the fermionic sectors (plus neurtinos and radiation) would lead to a proton to neutron ratio of 1:1. So the question is, whether in the scenario discussed here, the universe can get hot enough, such that it leads to full neutrino coupling (in the first generation of fermions), for a sufficient amount of time. Hence the number of protons and neutrons emerging in the right ratio along standard considerations when the universe cools due to the subsequent expansion.
\end{question}

The second major question is the question of whether the third generation (anti-)neutrino mass-eigenstates will in fact remain un-thermalized in the reheating process.
\begin{question}{(Missing) Thermalization of Tau Neutrinos }
At first glance the assumption that the third generation (anti-)neutrino mass-eigenstates remain largely unaffected by the reheating process seems plausible to me. The fermions of their own generation which they are most likely to interact with should be at low kinetic energies and it seems plausible that the reheating is too quick for neutrino oscillations to play a significant role before the neutrino decoupling time. Here the result in \cite{samuel1993neutrino} might be important, as it states that neutrino oscillations occur at a different rate in a dense neutrino gas. At least intuitively this seems to be the situation in the phase we are interested here.
\end{question}

Further what I expect to be a rather minor question.
\begin{question}{(Hot) Muon and Electron Neutrinos}
So far I have swept the term $\rho_{\nu_\mu, \nu_e}$  under the rug. This term describes the energy density of the (hot) muon and electron neutrinos which are generated in the reheating process. If this term would turn out to be of significant magnitude it would have two effects: 1) it would shift the radiation to mass energy ratio b) it would constitute a significant (hot) fraction of the Dark Matter sector. 
\end{question}
And last but not lest the elephant in the room:
\begin{question}{Can Neutrinos constitute Dark Matter}
There are two constraints on the role of Neutrinos as Dark Matter in current observations. One originates form CMB observations and the constraints on hot dark matter obtained there from. I believe that those bounds can not be applied to the here presented scenario based on the following considerations: \\
For this to work I need to assume that the third generation (anti-)neutrino mass-eigenstates will indeed pass the phase of reheating largely unaffected and will not be thermalized. This implies, that the third generation (anti-)neutrino mass-eigenstates remain in the kinetic state they were created in. Hence they remain approximately in a Fermi ground-state distribution where the particles are in the configuration with the lowest possible kinetic energy allowed by the Pauli principle. Hence the $\eps N(m_3, \Delta_{m_3}, \mathcal{A}, \Lambda,t_{\nu } )\alpha m_{\nu_3}$ part of $\rho_M(t_{\nu })$ would constitute ultra cold Dark Matter while only the neutrinos that make up $\rho_{\nu_\mu, \nu_e}$ constitute hot dark matter. Hence bounds on hot Dark Matter do not constrain the contribution of the third generation (anti-)neutrino mass-eigenstates in the scenario presented here.\\
The second constraint originates from the Gunn-Tremaine bound for fermionic Dark Matter in galaxies\cite{tremaine1979dynamical}. It rules out neutral leptons with a mass $m_{lepton}\lesssim 1MeV$ as dark matter in galaxies based on an argument regarding the available phase space volume and the Pauli principle. Current direct mass bounds on the third generation neutrino mass which I found in the literature are $m_{\nu_3}\leq 15.5MeV$. Which in principle still allows for third generation (anti-)neutrino mass-eigenstates to constitute Dark Matter in the galactic setting. Particularly since the third generation (anti-)neutrino mass-eigenstates would still be approximately in a Fermi ground-state for the galaxy (much similar to the electrons being in a Fermi ground-state when a star reaches the Chandrashekar limit)\footnote{This might actually have consequences for the initial evolution of supermassive black holes. However this has to be left to future work.}. However there are indirect limits from neutrino oscillations that place a much lower bound on $m_{\nu_3}$, which is well below the Gunn-Tremaine bound. There is one potential way to evade these indirect limits. In the scenario described here, we always observe neutrino oscillations in a dense neutrino gas and hence the considerations in \cite{samuel1993neutrino} might play a role and the effective neutrino oscillations would be substantially altered from vacuum neutrino oscillations. Hence the indirect bounds might be too low.
\end{question}

What is feeding into all these questions is the fact that at present I completely neglected the particles created in the first and second generation of fermions. This has to be kept in mind when thinking about the questions above. \\
In the following section I will discuss how the two previous sections can be fit together into one coherent story line for the evolution of the universe. 


\section{The Story Line} \label{42:sec:story}
\begin{figure}[h!]
    \centering
    \includegraphics[width=0.8\textwidth]{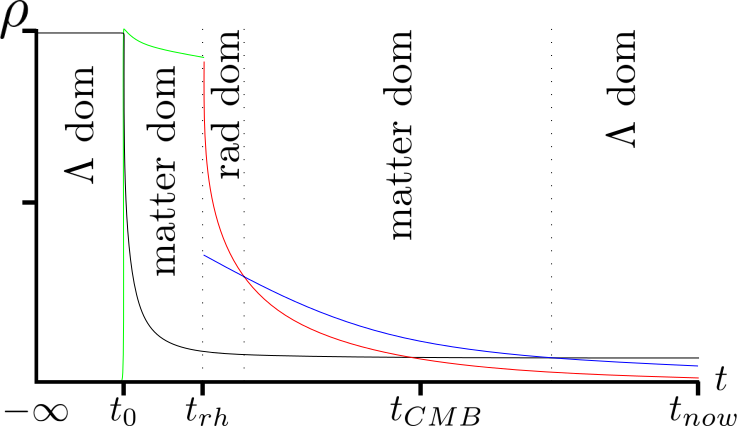}
    \caption{The graph shows a sketch of the expected development in "time" (horizontal axis) of the energy densities (vertical axis) associated with radiation (red), matter (blue) and $\Lambda$ (black) under the Friedman equations with a running vacuum model \eqref{42:eq:rchlambda}. The creation of matter in the third fermion generation and its evolution between the instability time $t_0$ and the reheating time $t_{rh}$ is sketched in green. I decided to use a seperate color due to all the uncertainties associated with that phase. Also because it breaks up into matter and radiation at reheating time. The universe starts of in a $\Lambda_{high}$ dominated phase, goes over into a short matter dominated phase right after the instability then becomes radiation dominated after reheating. Eventually it transitions to matter dominated and back to $\Lambda_{low}$ dominated as in the usual picture. I used $t_{now}$ to roughly indicate where in the evolution of the universe we are today and $t_{CMB}$ to roughly indicate where the CMB decoupled in this picture. }
    \label{42:fig:story1}
\end{figure}

In this section I will present a possible story line for the universe based on the discussions in section \ref{42:sec:lambda}, \ref{42:sec:asym} and \ref{42:sec:reheat}. In \cite{padmanabhan2017we} Padmanabhan argues that an arrow of time can emerge in a system with time symmetric equations if a suitable instability is present. This story line here is inspired by the ideas in the series of papers by Padmanabhan which was started with \cite{padmanabhan2012physical}. Most importantly I will consider the history of the universe to be that of a system going from a high energy (i.e. $\Lambda=\Lambda_{high}$) to a low energy (i.e. $\Lambda=\Lambda_{low}$) state. In contrary to the work of Padmanabhan however, I have the preliminary form of a mechanism that can describe the why and the how of this transition. In section \ref{42:sec:lambda} I discussed the fact that de-Sitter space is a stationary vacuum solution to the modified equations. Now if we allow for fluctuations in the metric and bring in the considerations from section \ref{42:sec:asym} and \ref{42:sec:reheat} one can see that the de-Sitter space with $\Lambda_{high}$ becomes unstable. The key to the instability lies in equation \eqref{42:eq:lambdader}. Before the fluctuations we have $\Lambda=\Lambda_{high}$ and $\rho_M=\rho_R=0$\footnote{It is important to realize that this period can in principle go on forever back in time. In fact if one considers time to be an emergent phenomenon describing a rate of change, such as was advocated by many discussants at the workshop "Progress and Visions in Quantum Theory in View of Gravity" at the Max Planck for Mathematics in Science in Leipzig in October 2018, then in this phase the concept of time ceases to make sense, because there is no such thing as change. Note that "past incompleteness" results such as \cite{borde2001inflationary} typically rely on geodesics, which implies the existence of a test particle. What I suggest here is that everything, all matter, radiation and even vacuum fluctuations have a creation time, i.e. their past directed trajectory is de facto incomplete. For radiation it is the reheating time, for particles the creation time and for fluctuations the point when they shrink below the Planck length. Let us extrapolate back in time. Let us consider the last fluctuation outgrowing the interacting region before creation time. We can trace that back to when it was smaller than Planck scale, hence when it disappeared. Nothing that happened before that moment plays any role for our universe today. One can therefore consider the pure, empty de Sitter space as an idealized past boundary condition. With the spacetime completely void of any structure an external observer could not tell the passage of time for the physical system she is looking at. On the level of the physical system any result that depends on a notion of test particles, i.e. geodesics, has no meaning if we assume the spacetime to be truly empty de Sitter space.}. As soon as the fluctuations set in the mass density $\rho_{M, 3g}$ of fermions in the third generation becomes non-vanishing and through the Friedmann equation \eqref{42:eq:friedmann} this leads to a change in the Hubble rate and a change in the radius $r_H$ of the causally interacting region. As soon as this leads to a decrease in the vacuum energy density more rapid than all other components on the right hand side of \eqref{42:eq:friedmann}, i.e. if
\begin{equation}\label{42:eq:instability}
\frac{d\Lambda}{d a}=-\frac{6}{r_{CH}^3}\frac{d r_{CH}}{da}\lesssim -\frac{4}{a^5},
\end{equation}
the instability is triggered. In particular the vacuum energy density has to decay quicker than the matter density $\rho_M \sim a^{-3}$ and the radiation density $\rho_R\sim a^{-4}$. Hence the system becomes matter or radiation dominated which increases $\frac{d r_{CH}}{da}$ substantially, decreasing the value of $\Lambda$ even faster. The influence of the vacuum energy then becomes rapidly negligible and the universe goes from a $\Lambda_{high}$ dominated phase to a matter/radiation dominated phase. This process continues until the radius of the interacting region has expanded so far, that the factor $1/r_{CH}^3$ becomes dominant and slows the decay of $\Lambda$\footnote{The future asymptotic behaviour is content of conjecture \ref{42:con:tods}.}. Once the instability is triggered the matter content $\rho_{M, 3g}$ decays as $1/a^3$ until reheating time $t=t_{rh}$. As before we assume instant reheating for simplicity. After $t_{rh}$ the radiation content and matter content evolve according to the standart picture with slight modifications due to the residual dynamic of $\Lambda$. Eventually matter and radiation disperse and we are left with a de Sitter vacuum again albeit at much lower value of $\Lambda=\Lambda_{low}$. A schematic representation of the evolution of the different energy components in the universe for this scenario can be found in figure  \ref{42:fig:story1}.\\
Note, although the scenario is inspired by the work of Padmanabhan, it is not reliant on it. In particular it is not clear, whether the here-presented scenario would lead to the "Cosmic Information" requirement,
\begin{equation}
    I_c= \frac{1}{9\pi}\ln{\left(\frac{4}{27}\frac{\rho_{\Lambda_{high}}^{3/2}}{\rho_{\Lambda_{low}}\rho_{eq}^{1/2} }\right)} =4\pi
\end{equation}
postulated by Padmamabhan, to be satisfied. In general it would be nice to have some sort of conservation law that links the different phases of the evolution of the universe\footnote{It would be interesting to see whether the set of idea developed by Padmamabhan \cite{padmanabhan2017cosmic, padmanabhan2014cosmological, padmanabhan2012physical, padmanabhan2017we, padmanabhan2016atoms, padmanabhan2014general, padmanabhan2016exploring, padmanabhan2017atoms, padmanabhan2015gravity} can be linked in to the discussion presented here. That would allow to interpret $I_c$ as a sort of conservation law for the transition. For this to be possible however a necessary requirement would be that $I_c=4\pi$ is true in the limit where the difference between the two energy levels goes to zero i.e. $\Lambda_{high}=\Lambda_{low}$, hence de Sitter space.}.
The following questions need further consideration:
\begin{question}{Dynamical $\Lambda$ and Nucleosynthesis}
After the instability is triggered the value of $\Lambda$ decreases rapidly. However no matter how quick that happens it might still have a sizable effect on the processes in the early universe. This phase between the instability and the CMB time in particular also contains the phase of nucleosynthesis. Whether and how this process would be altered and whether it would still be compatible with observations, remains an open question.
\end{question}

\begin{question}{Initial Condition}
What fixes the value of $\Lambda_{high}$?
\end{question}

\begin{question}{Trigger of instability}
The scenario discussed above ignores two crucial point: 
\begin{itemize}
    \item Why does the instability trace out a geodesic hypersurface (and thus lead to an observable Friedman reference frame in the universe)?
    \item How does the instability trigger over a large enough region simultaneously (note that this involves regions that are non-interacting at the time)
\end{itemize}
Essentially these are the questions associated with any inflationary scenario. Here of course the assumption to have no fluctuations until the instability is triggered makes no sense. But that condition was just introduced for the sake of the argument. In practice a detailed study of the interplay between fluctuations, matter creation and the trigger of the instability would be necessary.
\end{question}

On top of the questions mentioned here are of course all the ones mentioned in the sections before. 

\subsection{Slow Roll to Instability}
The problem regarding the initial conditions might be circumnavigated by taking into considerations that an empty de-Sitter space without fluctuations can never exist. Hence having the initial $\Lambda_{high}$ state sitting around for an arbitrary amount of time before the fluctuations kick in is unrealistic. What one could assume instead is that we start of with $\Lambda= 3/L_P^2$ and hence $r_{CH}=L_P$\footnote{It seems to me that this would ultimately be the highest possible value for $\Lambda$ that makes any physical sense.}. In the beginning, any matter that is created is dispersed so fast that it is almost neglectable on the right hand side of \eqref{42:eq:friedmann}. This leaves $\frac{d r_{CH}}{da}$ so small that despite $r_{CH}$ being very small the instability condition \eqref{42:eq:instability} is not satisfied. Therefore the value of $\Lambda$ decreases slowly until it reaches a value small enough (albeit still very high), that the matter creation can exceed the dispersive effect and the instability is triggered. 

\subsection{Dynamical Gravitational Coupling}
\begin{figure}[htb]
    \centering
    \includegraphics[width=0.75\textwidth]{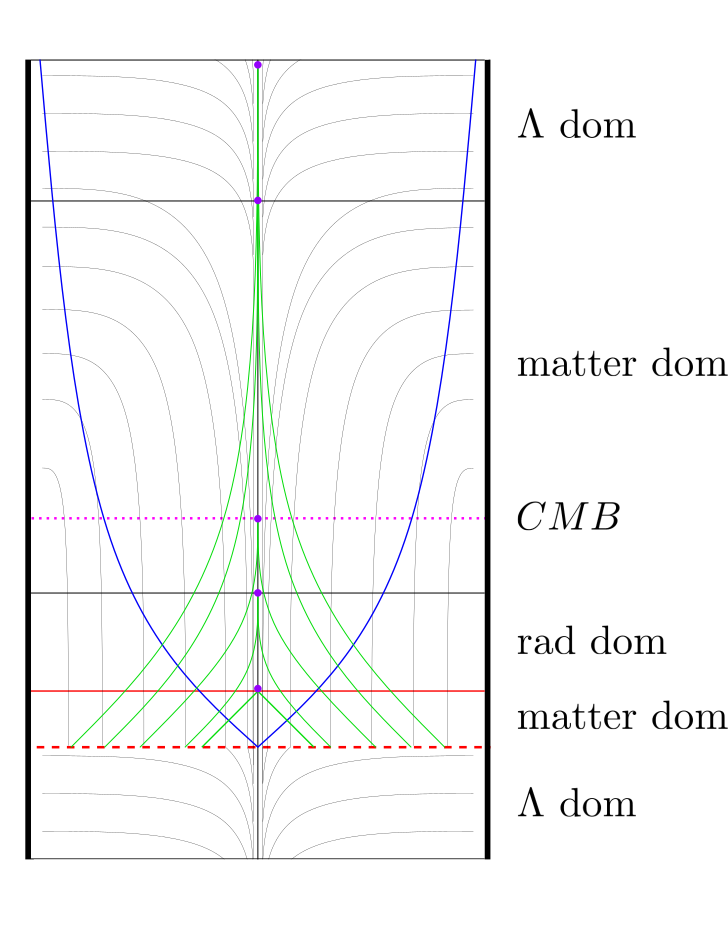}
    \caption{Here you see a schematic picture of the evolution of the universe for the scenario with a dynamical gravitational coupling. The evolution starts at the bottom and ends at the top. The sketch shows the region around an observer living on the central world line. The fine black lines indicate the geodesic spread in the spacetime. The thick black lines on the left and the right mark the boundary of the interacting region to the past and the future. The red dashed line indicates the matter creation time $t_0$. The blue line depicts the boundary of the interacting domain for matter created on the central world line at $t_0$. The green line depicts the backwards light cone of the observer on the central world line at different times (purple dots). The pink dotted line indicates the rough location of the CMB.\protect\\
    From the bottom up there is the first $\protect\Lambda$ dominated phase ($G=G_{low}$) then we have the creation (or "Big Bang") time $t_0$. This is followed by a short matter dominated phase before reheating. After that we have the normal sequence of radiation dominated-, matter dominated-, $\protect\Lambda$ dominated evolution ($G=G_{high}$) as in standard cosmology.\protect\\
    The sketch is intended to help the reader gain an intuition why the observer on the central world line can never see beyond the region that was causally interacting prior to $t_0$. Further the rough sketch for the interacting domain for matter shows how the standard FRW picture fits in. }
    \label{42:fig:dgchistory}
\end{figure}

A slightly different scenario occurs when we consider a dynamical gravitational constant (and hence a dynamical regularization) and a fix cosmological constant. In this case we would end up with a transition from a de Sitter vacuum with a small gravitational constant (i.e. small regularization length) to a de-Sitter vacuum with a large gravitational constant (i.e. a large regularization length). The picture should be again similar to before: 
\begin{itemize}
    \item De Sitter vacuum ($G=G_{low}$)
    \item Fluctuations create matter
    \item Instability triggered
    \item (almost) FRW evolution
    \item back to De-Sitter vacuum ($G=G_{high}$)
\end{itemize}
Now there is something particular about the initial and final state: As they are void of any matter or radiation the gravitational constant doesn't matter and since we chose $\Lambda$ to be constant we arrive at the conclusion that the physical size of the interacting region in the initial and the final state is identical. But wait, don't we know that in FRW the size of the interacting region increases? So how can the physical size of the interacting region be the same in the beginning and the end? The following explanation seems plausible: For the matter interaction the instability time slice at creation time $t=t_0$ acts as an initial data surface. Therefore the horizon for the matter interaction vanishes at that time and expands from there. The vacuum fluctuations prior to $t_0$ are correlated across the entire interacting region and imprint on the matter distribution at creation time. However the matter interaction only starts once matter is created. In a sense that merges the Big Bang with the inflationary picture. As discussed before: If we assume the absence of fluctuations prior to $t_0$ then in empty de Sitter space the notion of time becomes void of meaning. Physical time as a measurement of the rate of change comes into existence when the fluctuations start and matter is created. Of course vacuum fluctuations always exist and they trace out the pre $t_0$ "inflationary" de Sitter phase. \\
This resolves two of the conceptional issues from the dynamical $\Lambda$ consideration:
\begin{itemize}
    \item The size of the interacting region before the instability is identical to the size in the asymptotic future. That means, that an observer can never see any region beyond what was already interacting before the instability triggered. Therefore it is natural that the entire region of the CMB that we can see is correlated because it was in fact one interacting region before the instability triggered. Hence the vacuum fluctuations would naturally be correlated across the entire region.
    \item It is sufficient (and plausible) for the instability to be triggered simultaneously across the interacting region. Hence across the maximal domain that we will ever be able to observe. 
    \item It seems conceivable that the vacuum fluctuations would be "homogenous" across the interacting region, and would hence trace out a geodesic hypersurface. Therefore the instability would also trace out a geodesic hypersurface, essentially giving us the FRW reference frame that we observe for the CMB.  
\end{itemize}
See Figure \ref{42:fig:dgchistory} for an illustration of this scenario. 

\begin{figure}[htb]
    \centering
    \includegraphics[width=0.8\textwidth]{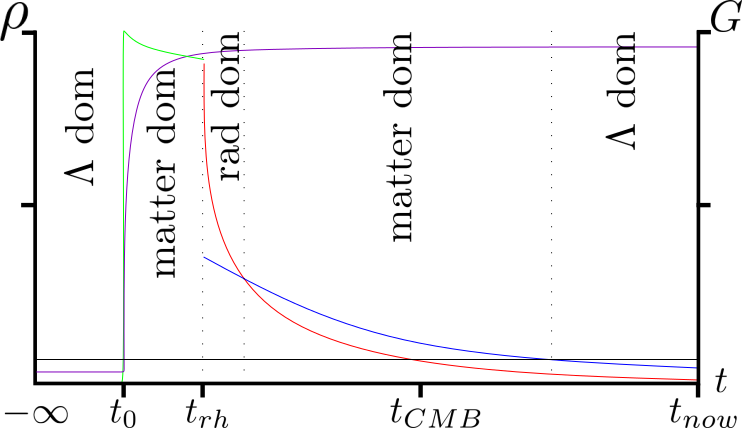}
    \caption{The graph shows a sketch of the expected development in "time" (horizontal axis) of the energy densities (vertical axis) associated with radiation (red), matter (blue) and $\Lambda$ (black) under the Friedman equations with a dynamical gravitational constant (second vertical axis on the right). The evolution of the gravitational constant is freely made up to fit the picture and is here drawn in purple. The creation of matter in the third fermion generation and its evolution between the instability time $t_0$ and the reheating time $t_{rh}$ is sketched in green. I decided to use a separate color due to all the uncertainties associated with that phase. Also because it breaks up into matter and radiation at reheating time. The universe starts of in a $\Lambda$ dominated phase with a small gravitational constant $G$, goes over into a short matter dominated phase right after the instability then becomes radiation dominated after reheating. Eventually it transitions to matter dominated and back to $\Lambda$ dominated as in the usual picture but now with a gravitational constant significantly bigger than in the initial $\Lambda$ dominated phase. I used $t_{now}$ to roughly indicate where in the evolution of the universe we are today and $t_{CMB}$ to roughly indicate where the CMB decoupled in this picture.}
    \label{42:fig:story2}
\end{figure}

One observation though that is somewhat worrisome is that the difference in the Gravitational constant in the beginning and the end, i.e. the different regularizations $L_{P,begin}<L_{P,end}$, imply
\begin{equation}
S_{C, begin}>S_{C, end}.
\end{equation}
This can be seen as the gravitational degrees of freedom being transferred to the matter degrees of freedom\footnote{At the moment this is in fact my best guess for how the instability mechanism could work in this scenario.}. Those degrees of freedom then eventually leave the interacting region and thus carry away the degrees of freedom associated with them. A rough sketch of this scenario for the evolution of the universe is given in figure \ref{42:fig:story2}. \\
As a remark on the side, it is interesting to note, that with a fix value of $\Lambda$ we get in a sense a fix physical domain that is sufficient to describe everything we will ever be able to observe. Together with a finite regularization this should leave us with a finite volume, finite Hilbert space CFS. In this setting the variational problem \eqref{42:eq:actionprinciple} is known to be well-posed. \\
The following questions need further consideration: 
\begin{question}{Instability Mechanism}
The biggest open question in this scenario is the question, how the instability mechanism could be triggered. It is plausible, that the same mechanism that leads to matter creation in the CFS framework, also leads to a change in the regularization and hence in Newtons coupling constant.
\end{question}

\begin{question}{Suppression of Particle Creation}
As the matter creation mechanism comes in with the regularization length scale the dynamical gravitational coupling with increasing $G$ (i.e. dynamical regularization) would in the present form enhances the effect as time progresses. This question needs careful investigation as this would obviously lead to problems with observations. 
\end{question}

\section{Discussion of the Causal Fermion Systems Framework}\label{42:sec:cfsdisc}
Given the little consideration the CFS framework has received in the literature so far, I feel compelled to include at this point a discussion section of the most interesting results derived from it. This serves to demonstrate that the considerations in the present paper are part of a much larger picture. I will also include some new view points and how the framework might be linked to other considerations. Lets us first introduce a distance function along the lines of \cite{distancefinster}:
\begin{definition}[Distance Function]
We will refer to $l(x,y)^2=(|xy|)^{-1/3}$ as the distance between two points. 
\end{definition}
Here $|xy|=\sum_{i=1}^4 |\lambda^{xy}_i|$ is just the sum of the absolute values of the eigenvalues of the operator $xy$. Note that $|xy|$ has dimension $[length^{-6}]$ and therefore $(|xy|)^{-1/3}$ has dimension $[length^2]$. The main motivation to introduce this distance function is to establish another relation to the considerations in \cite{padmanabhan2017atoms}. Note that in the CFS framework a point $x$ is timelike separated from itself. However taking definition \ref{42:def:order} serious it lies neither in its own future nor its own past because by definition ${\mathscr{C}}(x,x)=0$ and thus we have no time direction. However one expects the following to hold
\begin{conjecture}[Zero Point Distance]
In any minimizer we have that $l(x,x)^2\sim\epsilon^2$.
\end{conjecture}
Thereby the distance function would satisfy the modification made by Padmanabhan on the geodesic distance function when deriving the mesoscopic effects of internal degrees of freedom of gravity (i.e. mesoscopic effects of quantum gravity).\\
 Alternatively one could consider the following causal distance function: $\sigma(x,y)= ({\mathcal{L}}[A_{xy}])^{-2/3}$. This is just a simple redefinition of the Lagrangian \eqref{42:eq:actionprinciple} but now one can say that instead of minimizing the amount of correlation within a spacetime, minimizing the action, maximizes the point wise causal distance in the spacetime. The causal distance function has the property that $\sigma(x,y)<\infty$ for $x,y$ causally separated and $\sigma(x,y)=\infty$ for $x,y$ spacelike separated. The action can then be stated as:
\begin{equation}
    {{\mathcal{S}}} = \iint_{\F \times \F} \sigma(x,y)^{-3/2}\: d\rho(x)\: d\rho(y).
\end{equation}
The link to Padmanabhans considerations would then be less clear.\\
Another set of thoughts that one might try to relate to the CFS formalism, is the Events, Trees, Histories (ETH) interpretation of Quantum Mechanics, developed and propagated by Fr\"ohlich \cite{frohlich2016quest}. The intuitive approach would be to interpret $\F$ as the set of all possible physical events over a Hilbert space. Correspondingly the support of the measure $M$ would then identify those events that are actually realized in a physical system. We are currently investigating whether these two frameworks can indeed be connected in a meaningful way. It seems in general that the CFS framework would be compatible with any sort of relational interpretation of quantum mechanics. 

\subsection{Relevant Results}\label{42:sec:results}
I will here just state some of the results derived from the CFS framework that are the most interesting ones from a physical point of view in my opinion. They can be found in \cite{finster1542continuum}. 
\begin{itemize}
\item Emergent spacetime and bosonic fields including field equations. 
    \item The CFS framework works only if there are at least 3 generations of fermions. 
    \item The gauge group of the Standard Model as well as the mixing matrices can be derived\footnote{It is unclear how rigid the framework is with that respect. For example, it is unclear whether one could accommodate for axions, by adding additional fermionic sectors.}.
    \item The emergent spacetime has exactly one timelike direction. The maximum extension of the Clifford Algebra is 5 dimensional (i.e. if we reserve one dimension for chirality there are at maximum 3 spacelike dimensions). 
    \item When mapping a spacetime with fields into the CFS framework, takeing a unitary transformation $U$ of the resulting measure $\rho$ gives us another minimizer $\tilde\rho$. A convex combination then has a smaller action. This is referred to as microlocal mixing. This procedure has an effective description in spacetime as a Fock space, i.e. as the second quantized description of the fields in the spacetime. In that sense, second quantization seems to emerge from the minimizing of the causal action principle!
    \item The vacuum polarization graph comes out finite with the identical value as from renormalization. 
\end{itemize}
Given the complexity of the CFS framework and the small number of people proficient in it, I would advice a mild amount of precaution despite all these claims being backed up by peer reviewed papers. 

\subsection{Open Questions}
I will now mention three of the serious open questions that I am aware of regarding the CFS framework.

\begin{question}{Missing Higgs}
The Higgs field has not yet been worked out. According to Finster, however this is work in progress and looks promising.
\end{question}

\begin{question}{What does it tell you?}
So far from my point of view this is one of the biggest questions I have with regard to the framework. Its central object i.e. the action principle including all constraints, has no simple physical interpretation. 
\end{question}

\begin{question}{What is necessary for the construction?}
It is unclear, which is the minimal set of observational facts, that one has to build into the framework by hand, to recover everything we know and love about the universe.  
\end{question}

\section{Conclusion}\label{42:sec:conclusion}
In the present work I discussed how the matter creation mechanism derived from the CFS framework paired with an adequate modification of gravity can lead to a coherent story line for the universe. Most importantly I discussed how the matter creation mechanism allows for Cold Dark Matter to be made up of third generation (anti-)neutrino mass-eigenstates with minimal kinetic energy. At this point I want to stress how essential the last point is. Despite the many gaps in the argument it is clear that the proposed Dark Matter candidate enters with the right kinematic.\\
On top of that, the scenarios presented here would also resolve the problem of the matter/anti-matter asymmetry. Further more the scenario suggests a new angle on the issue of the initial ``singularity'' and the cosmological constant.\\
Given, that the Cold Dark Matter candidate is within the Standard Model, I would expect this prediction to be testable with current technology or maybe even currently already available data. In the following list I collect some results that might be worth checking against this new prediction. 
\begin{itemize}
    \item A residual creation of 3rd generation fermions and their decay 
could create a distinct fingerprint of high energy radiation in the late time univers.
\item Can the observed annual modulation in events in the DAMA/LIBRA detector \cite{bernabei2013dama} be explained with third generation (anti-)neutrino mass-eigenstates? The DAMA/LIBRA experiment observes an annual modulation in events below the $6KeV$ level in a NaI(Tl) scintillation detectors. This modulation is consistent with the signal one would expect based on the changing relative velocity of earth with respect to the galactic halo as it orbits the sun.   
\item Does the chirality asymmetry in the Dirac sea or in the dense third generation (anti-)neutrino mass-eigenstate background have an effect on the results of a pear shaped nucleus reported in \cite{bucher2016direct}?
\item One very interesting experimental result is the $B_s^0$ and $B^0$ Meson decay into muon pairs investigates at the ATLAS Experiment at CERN \cite{atlas2016study}. This result is particularly interesting, as the mesons in question are made up of a bottom quark and a lighter partner. Hence an interaction with the dense third generation (anti-)neutrino mass-eigenstate background might play a role.
\item In \cite{krasznahorkay2016observation} a slight deviation from the Standard Model predictions is reported in an nuclei decay. They calculate the mass of the involved additional particle to be around 17 MeV which is roughly in the ball park of the bounds for $m_{\nu_3}$. However if I understand right they suggest a bosonic particle of that mass, which would of course be incompatible with an interaction with neutrinos.
\item Neutrino experiments such as T2K \cite{batkiewicz2017latest} and MiniBooNE \cite{aguilar2018observation} might be sensitive to a dense third generation (anti-)neutrino mass-eigenstate background. 
\item Observations show different lifetimes for neutrons whether they are kept in a bottle \cite{pattie2017measurement} or measured in a beam. Though it doesn't seem very likely that this were connected to a dense third generation (anti-)neutrino mass-eigenstate background. 
\item The absorption spectrum detected by EDGES \cite{bowman2018absorption} suggests that the gas in the early universe was colder than previously assumed. The fact, that the third generation (anti-)neutrino mass-eigenstate background in the here-presented scenario is created with minimal kinetic energy, might explain this observation if the gas in the early universe was able to loose enough kinetic energy to the cold third generation (anti-)neutrino mass-eigenstate background, i.e. if the third generation (anti-)neutrino mass-eigenstates can act as a heat sink. An important question is whether the all the  background third generation (anti-)neutrino mass-eigenstates can exchange kinetic energy with the gas, or only the fraction that, due to neutrino oscillation, is in the electron neutrino weak interaction state. 
\item It is unclear how a dense gas of fermionic particles effects early gravitational collapse in general and the growth rate of the first black holes in collapsing gas clouds specifically.
\end{itemize}
These are all experiments and observations that I am aware of, that might be affected when taking a dense third generation (anti-)neutrino mass-eigenstate background into consideration\footnote{I will discuss some of the consequences for some of these experiments in more detail in an upcoming paper.}. There might of course also be some experiments that fit Standard Model predictions perfectly well, that would be affected and could help to falsify the prediction of a dense third generation (anti-)neutrino mass-eigenstate background.\\
As already discussed in the introduction the paper does not produce a complete theory, but it should be considered as a backbone for further investigations. The 15 open questions mentioned through out the paper hopefully give a comprehensive picture of the gaps that remain to be filled.\\
To conclude I think it is needless to say that a lot more research is necessary to put the ideas presented here on firm theoretical ground. 

%
\begin{acknowledgement}
I would like to thank Felix Finster, Todd Oliynyk, Emmanuel Saridakis, Erik Curiel, Calum Robertson, Markus Strehlau, Mark Bugden, Marius Oancea and Isha Kotecha for listening patiently to my clumsy explanations in the early stage of the development of these ideas. I would like to thank Alice Di Tucci for explaining to me the debate surrounding the initial singularity and pointing me to relevant sources. \\
I would like to thank Thanu Padmanabhan and Hamsa Padmanabhan for helpful discussions during my visit in Zurich. \\
This work was supported by the Australian Research Council grant DP170100630.\\
I am indebted to Ann Nelson, who brought the problem with the Gunn-Tremaine bound to my attention.
\end{acknowledgement}
%

%
%
%

%
%
\bibliographystyle{plain}
\bibliography{proposal42.bib}

\end{document}